\documentclass[%
 twocolumn,
 amsmath,amssymb,
 aps,
 prl,
]{revtex4-1}

\usepackage{graphicx}
\usepackage{dcolumn}
\usepackage{bm}
\usepackage[colorlinks]{hyperref}

\renewcommand{\vec}[1]{{\bf #1}}
\newcommand{\braket}[1]{\langle #1  \rangle} 
\newcommand{\DI}{\text{DI}} 
\newcommand{\nn}{\text{nn}} 
\newcommand{\nnn}{\text{nnn}} 
\newcommand{\M}{\text{M}} 
 
\newcommand{\RFIM}{\text{RFIM}} 
\newcommand{\TFIM}{\text{TFIM}} 
\newcommand{\Trotter}{\text{Trotter}} 

\begin{document}


\title{Monte Carlo Renormalization Group for Classical Lattice Models with Quenched Disorder}

\author{Yantao Wu$^1$ and Roberto Car$^{1, 2}$}

\affiliation{%
$^1$The Department of Physics, Princeton University\\
$^2$The Department of Chemistry, Princeton University \\
}%


\begin{abstract}
We extend to quenched disordered systems the variational scheme for real space renormalization group calculations that we recently introduced for homogeneous spin Hamiltonians. 
When disorder is present our approach gives access to the flow of the renormalized Hamiltonian distribution, from which one can compute the critical exponents if the correlations of the renormalized couplings retain finite range.
Key to the variational approach is the bias potential found by minimizing a convex functional in statistical mechanics. 
This potential reduces dramatically the Monte Carlo relaxation time in large disordered systems. 
We demonstrate the method with applications to the two-dimensional dilute Ising model, the random transverse field quantum Ising chain, and the random field Ising in two and three dimensional lattices.
\end{abstract}

\pacs{Valid PACS appear here}
\maketitle


Understanding the phase diagram of quench-disordered systems, such as glasses or materials with a disordered distribution of defects, is a major scientific goal. 
The effect of fluctuations on the equilibrium properties of translationally invariant spin models has been studied successfully with real space Monte Carlo (MC) renormalization group (RG) techniques \cite{mcrg,mcrg_rc, iMCRG}, but direct MCRG studies of disordered systems face major difficulties.  
In random systems, the RG flow of the Hamiltonian distribution is of fundamental importance \cite{lubensky}. 
Its explicit calculation requires an average of the RG flows of many Hamiltonians, each with an extensive number of quench-disordered couplings. 
Moreover, in disordered systems MC relaxation times tend to be significantly longer than in pure systems.  
Although it is an old idea to study quench-disordered systems with real-space renormalization, the task is so computationally challenging that it has not been explicitly carried out within MCRG.
So far, the challenge of dealing with many random couplings has been avoided, either by limiting the form of the disordered renormalized Hamiltonian, or by adopting techniques that do not require its explicit calculation \cite{parisi, wang_sg}. 

Recently, we introduced a scheme called Variational Monte Carlo Renormalization Group (VMCRG) \cite{vmcrg} that facilitates the calculation of the renormalized coupling constants and critical exponents by mitigating the effects of critical slowing down.  
Here we show that this approach makes possible to compute directly the evolution of the coupling distribution under scale transformations in classical quench disordered models, in addition to greatly alleviating sampling difficulties due to disorder. 
The method is particularly useful when dealing with finite disorder fixed-points whose critical distribution has a finite width that is difficult to estimate perturbatively. 
In these situations, VMCRG recovers the scaling law for the singular part of the free energy, and leads to a viable scheme for computing the critical exponents, when the evolving distribution can be parameterized in terms of local correlations between the renormalized couplings. 
The approach can also discern strong disorder fixed-points characterized by a diverging variance of the critical distribution, but in this case, it does not provide a way to compute the critical exponents. 
Strong disorder fixed points have often been associated to disordered quantum models that are amenable to exact solution with Strong Disorder Renormalization Group (SDRG) techniques \cite{fisher_prl, fisher_prb}. 
If the partition function of these systems has a sign-free path integral representation, the corresponding classical model can be studied numerically with VMCRG, which then provides an alternative way of assessing the strong disorder character of the critical distribution.

We illustrate the formalism with applications to four disordered spin systems, namely, the 2D dilute Ising model (DIM), the Trotter approximation of the 1D random quantum transverse-field Ising model (TFIM), and the random field Ising model (RFIM) in 2D and 3D. 
We find the following results. 
The critical Hamiltonian distribution of the 2D DIM approaches a finite-disorder fixed point, indicating that disorder persists at all length scales, a condition that is difficult to establish with perturbative means as disorder is neither asymptotically small or large. 
We find that disorder has an even larger effect in the 1D TFIM, where the magnetic phase transition is associated to a Hamiltonian distribution with increasing variance along the RG flow, consistent with the strong-disorder fixed point predicted by SDRG \cite{fisher_prl, fisher_prb}. 
The magnetic phase transition is wiped out by disorder in the 2D RFIM, where we find that, well below the critical temperature of the pure 2D Ising model, the RG flow approaches the fixed-point with zero-field and zero-coupling, in agreement with the exact result for the free energy of this model \cite{aizenman}. 
In the 3D RFIM there is a magnetic phase transition, in agreement with earlier predictions \cite{3D_rfim}. 
The corresponding critical distribution, however, is not of the finite-disorder kind like in the 2D DIM, but shows increasing variance along the RG flow, like in the 1D TFIM. 
Interestingly, this behavior, i.e. an increasingly large variance without significant changes in the mean along the RG flow, is also observed within a finite coupling range below the critical coupling, suggesting that the random magnetic fields promote strong disorder in the RG distribution of this model even below bulk criticality.

In the following we consider a generic quench-disordered spin Hamiltonian with local interactions on a lattice of $N$ sites: 
\begin{equation}
\label{eq:hamiltonian} 
H_{\vec K} (\bm \sigma) = -\sum_{\alpha} \sum_{i =1}^N \sum_{s=1}^{N_s(\alpha)} K_{\alpha}^{i,s} S_\alpha^{i,s}(\bm \sigma), \hspace{2mm} \vec K \sim P_{\vec v}(\vec K) 
\end{equation}
Here the index $\alpha$ specifies the coupling type, such as nearest neighbor, smallest plaquette, etc. 
The index $i$ runs over the $N$ lattice sites, while $s$ runs over the $N_s(\alpha)$ point group symmetry operations that generate distinct couplings of type $\alpha$ stemming from site $i$. 
For example, the nearest neighbor coupling has two terms at each lattice site, while the smallest plaquette has only one.
$S_\alpha^{i,s}$ are products of spins in the neighborhood of $i$ specified by $\alpha$ and $s$.  
The coupling constants $K^{i,s}_\alpha$ are made dimensionless by incorporating the factor $(k_B T)^{-1}$ in their definition. 
The vector $\vec K$ denotes the full set $\{K_\alpha^{i,s}\}$ of couplings corresponding to a disorder realization drawn from the probability density $P_{\vec v}(\vec K)$ specified by the parameter set $\vec v$. 

Let $\bm\sigma' = \tau(\bm\sigma)$ be a coarse-graining map, such as the block spin transformation \cite{block2}, which implements a scale dilation that preserves the symmetry of $P_{\vec v}(\vec K)$.
The corresponding renormalized couplings $\vec K'$ and Hamiltonian $H'_{\vec K'}$ are 
\begin{equation}
  \label{eq:renormalized_h}
  H'_{\vec K'}(\bm\sigma') + Ng(\vec K)= -\ln \sum_{\bm\sigma} \delta_{\tau(\bm \sigma), \bm\sigma'} e^{-H_{\vec K}(\bm\sigma)} 
\end{equation}
Here $\delta_{\tau(\bm\sigma), \sigma'}$ is the Kroneker delta function. $g(\vec K)$ indicates the ``background'' free energy per site of a RG transformation \cite{nauenberg} so that $H'_{\vec K'}$ does not contain spin independent terms. 
Let $\mathcal{R}$ be the RG map of the coupling constants implicitly defined by Eq. \ref{eq:renormalized_h}:    
\begin{equation}
  \vec K' = \mathcal{R}(\vec K)
\end{equation}
The distribution of the renormalized constants $P_{\vec v'}(\vec K')$ is related to $P_{\vec v}(\vec K)$ by  
\begin{align}
  P_{\vec v'}(\vec K') &= \int d\vec K P_{\vec v}(\vec K) \delta(\vec K' - \mathcal{R}(\vec K)) 
\end{align}
Thus, the renormalization of the coupling constants, from $\vec K$ to $\vec K'$, induces a renormalization from $\vec v$ to $\vec v'$. 
In disordered systems, $\vec v$ plays the role of scaling variable. 
 
In our procedure we calculate $\vec K' = \mathcal{R}(\vec K)$ for a representative number of quenched realizations.  
Each map involves a large number of disordered coupling constants. 
Sampling is hampered by the rugged disordered energy landscape and is slowed down by long-range correlations near criticality. 
VMCRG overcomes these difficulties by adding to the renormalized Hamiltonian $H'_{\vec K'}(\bm\sigma')$ a bias potential $V(\bm\sigma')$ so that the distribution of $\bm\sigma'$ under the Hamiltonian $H'_{\vec K'}(\bm\sigma') + V(\bm\sigma')$ becomes equal to a preset target probability $p_t(\bm\sigma')$. 
By choosing the uniform distribution for the latter, i.e. $p_t(\bm\sigma') = (\frac{1}{2})^{N'}$ for Ising systems, the variables $\bm\sigma'$ are uncorrelated.   
Thus, finite size effects are greatly reduced because, in the biased system, the correlation functions decay exponentially over a distance approximately equal to $b$, the linear size of the block spin, even at criticality. 
Following \cite{varyfes} the bias potential that performs this task minimizes the convex functional $\Omega[V]$ given by:  
\begin{equation}
  \label{eq:Omega}
  \Omega [V] = \ln\frac{ \sum_{\bm \sigma'} e^{-[H'(\bm \sigma') + V(\bm \sigma')]}}{\sum_{\bm \sigma'} e^{-H'(\bm \sigma')}}+ \sum_{\bm \sigma'} p_t (\bm \sigma') V(\bm \sigma') 
\end{equation}
The minimizing potential, $V_\text{min}$, satisfies \cite{vmcrg}: 
\begin{equation}
  \label{eq:vmin}
  H'_{\vec K'}(\bm\sigma') = - V_{\text{min}}(\bm\sigma'), 
\end{equation}
modulo an immaterial constant. 
Thus, by minimizing $\Omega$ one finds the renormalized Hamiltonian. 
In practice, we adopt for $V$ a finite representation that parallels the one of the Hamiltonian in Eq. \ref{eq:hamiltonian}:  
\begin{equation}
  V_{\vec J}(\bm\sigma') = \sum_\alpha \sum_{i = 1}^N \sum_{s=1}^{N_s(\alpha)}J^i_\alpha S^i_\alpha(\bm\sigma') 
\end{equation} 
The minimizing coefficients, $\vec J_{\text{min}} = \{J_{\alpha, \text{min}}^i\}$, can be found by minimizing $\Omega$ with a gradient descent procedure \cite{varyfes}. 
In disordered systems, the number of unknown coefficients is large and we use only the diagonal part of the Hessian $\frac{\partial^2 \Omega}{\partial J_\alpha^{i,s}\partial J_\beta^{j,t}}$ in addition to the gradient $\frac{\partial \Omega}{\partial J_\alpha^{i,s}}$ in the minimization procedure. 
Empirically, we find that the optimization cost increases linearly with the number of coefficients $J_\alpha^{i,s}$, making possible calculations on large lattices. 
We have by virtue of Eq. \ref{eq:vmin}: 
\begin{equation}
  \vec K' = -\vec J_{\text{min}}
\end{equation}

The RG procedure is repeated for $N_D$ disorder realizations, generating many $\vec K'$ vectors distributed according to $P_{\vec v'}(\vec K')$ at each RG iteration.  
This distribution can be visualized with histograms representing the marginal distribution of coupling type $\alpha$: 
\begin{equation}
  \label{eq:histogram}
  Q_{\vec v}(K_\alpha) = \sum_{i_D=1}^{N_D} \sum_{i=1}^{N}\sum_{s = 1}^{N_s(\alpha)} \frac{\delta_\epsilon(K_\alpha - (K_\alpha^{i,s})_{i_D})
}{N_DN_\alpha N_s(\alpha)} \end{equation}
Here $\delta_\epsilon$ is a delta-function approximant with support $\epsilon$. 

When the critical fixed-point distribution has finite disorder the following procedure can be used to compute the critical exponents.  
We indicate by $\vec v^*$ the parameter set corresponding to the critical distribution.
In order to compute the critical exponents we should compute the leading eigenvalue(s) of the Jacobian of the transformation of the scaling variables, $\frac{\partial \vec v'}{\partial \vec v}$, at $\vec v^*$. 
Assuming that the correlations between the couplings are short ranged we may use a finite set of short-ranged basis functions $U_\beta(\vec K)$ to represent $P_{\vec v}(\vec K)$ \cite{parisi}: 
\begin{equation}
  \label{eq:log_p}
  -\ln P_{\vec v}(\vec K) = C + \sum_\beta v_\beta U_\beta(\vec K)
\end{equation}
Here $C$ is a normalizing constant and the index $\beta$ specifies the coupling correlation type, such as one-body, two-body, etc., associated to products of different $K_\alpha$ or combinations thereof. 
The sum over $\beta$ includes terms of increasing range up to some cutoff distance on the lattice. 
The vector parameter $\vec v$ corresponds to the set of amplitudes $\{v_\beta\}$. 
The coupling functions $U_\beta(\vec K)$ are sums of local coupling products that play a role similar to that of the spin functions $S_\alpha^{i,s}(\bm\sigma)$ in Eq. \ref{eq:hamiltonian}.  
For example, for Hamiltonians with nearest neighbor ($K_{\nn}$) and next nearest neighbor ($K_{\nnn}$) couplings, the first four $U_\beta(\vec K)$ could be $U_1 = \sum_{i, s} K_{\nn}^{i,s}$, $U_2 = \sum_{i,s} K_{\nnn}^{i,s}$, $U_3 = \sum_{i,s} (K_\nn^{i,s})^2$, and $U_4 = \sum_{i,s} K_{\nn}^{i,s}K_{\nnn}^{i,s}$.
Taking the derivative $\frac{\partial \vec v'}{\partial \vec v}$ in the close proximity of $\vec v^*$, we obtain: 
\begin{equation}
\label{eq:matrix_eq}
\braket{U_\beta U'_\gamma} - \braket{U_\beta}\braket{U'_\gamma} = \sum_\alpha \frac{\partial v'_\alpha}{\partial v_\beta} \cdot  (\braket{U'_\alpha U'_\gamma} - \braket{U'_\alpha}\braket{U'_\gamma}) 
\end{equation}
Here $\braket{\cdot}$ denotes an average under $P_{\vec v}(\vec K)$. 
The RG Jacobian may then be obtained from Eq. \ref{eq:matrix_eq}. 

We now consider systems on the square and cubic lattices to demonstrate the method. 
We start with a DIM, which in 2D is marginal for the Harris criterion \cite{harris} that is commonly used to characterize whether disorder is relevant at criticality. 
The Hamiltonian is 
\begin{equation}
  \label{eq:dilute}
  H_{\DI} = -K_{\DI}\sum_{\braket{i, j}} k^{ij} \sigma_i\sigma_j
\end{equation}
Here $K_{\DI} > 0$, $\braket{i, j}$ denotes nearest neighbors, and $k^{ij} = 1$ or $\frac{1}{2}$ with probability $\frac{1}{2}$.
The critical value of $K_{\DI}$ is known to be $K_{\DI,c} = 0.609377...$ by a duality argument \cite{di_duality}. 
We adopt the majority rule with a random tie-breaker on $b\times b$ blocks with $b = 2$. 
Three couplings are included in the renormalized Hamiltonian, namely nearest neighbor ($K_{\text{nn}}$), next nearest neighbor ($K_{\text{nnn}}$), and smallest plaquette ($K_{\square}$), which are the most important couplings in the pure Ising model. 
The calculations are done on $128^2$ lattices for 4 RG iterations for three values of $K_{\DI}$, i.e. $K_{\DI} = K_{\DI, c}, 0.60$, and 0.62. 
In addition, for $K_{\DI} = K_{\DI,c}$, we carry out a $5$th iteration on a $256^2$ lattice. 
For $n = 5$, we deal with spin blocks of linear size $b^n = 32$, for which spin correlations are significant. 
In this case, we find that sampling efficiency improves significantly by adopting the Wolff algorithm \cite{wolff} instead of the Metropolis algorithm \cite{metropolis} used in all other simulations in this paper. 
For the simulation correlation time, $\tau\sim \xi^z$, the cluster algorithm reduces the dynamical exponent $z$ while the bias potential reduces the correlation length $\xi$. 

\begin{figure}[hth]
\centering
\begin{minipage}{.155\textwidth}
  \centering
  \includegraphics[scale=0.19]{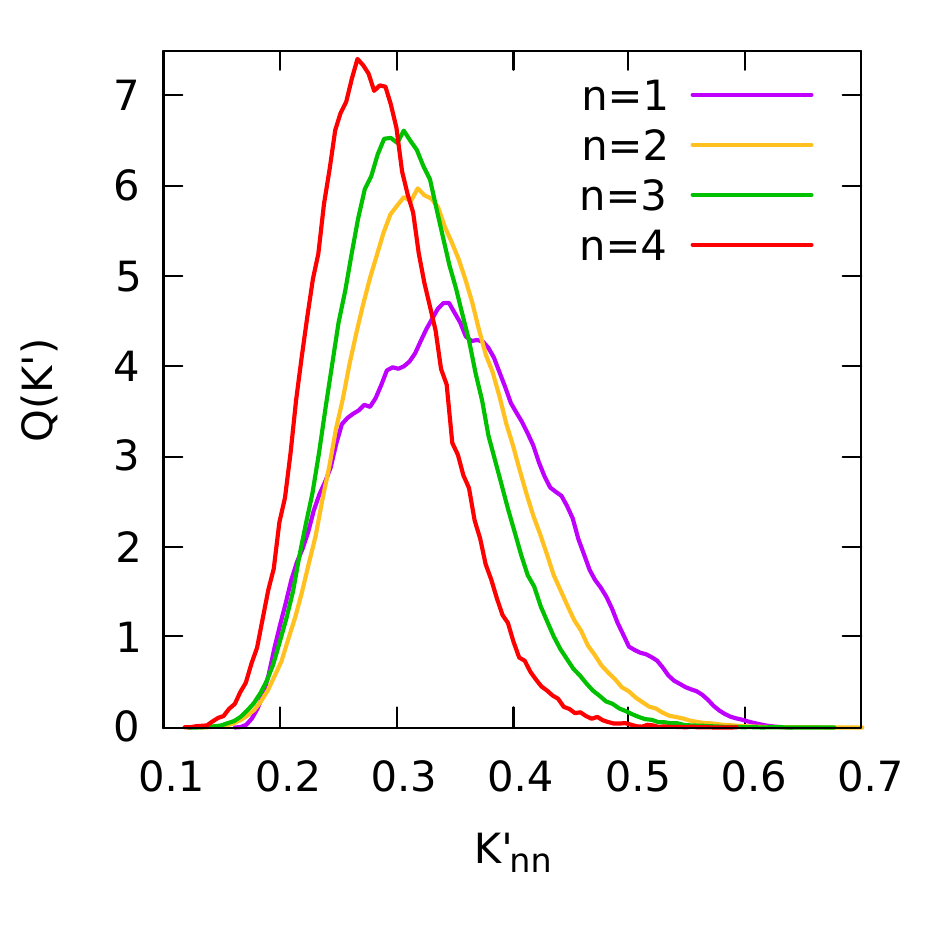}
\end{minipage}%
\begin{minipage}{.155\textwidth}
  \centering
  \includegraphics[scale=0.19]{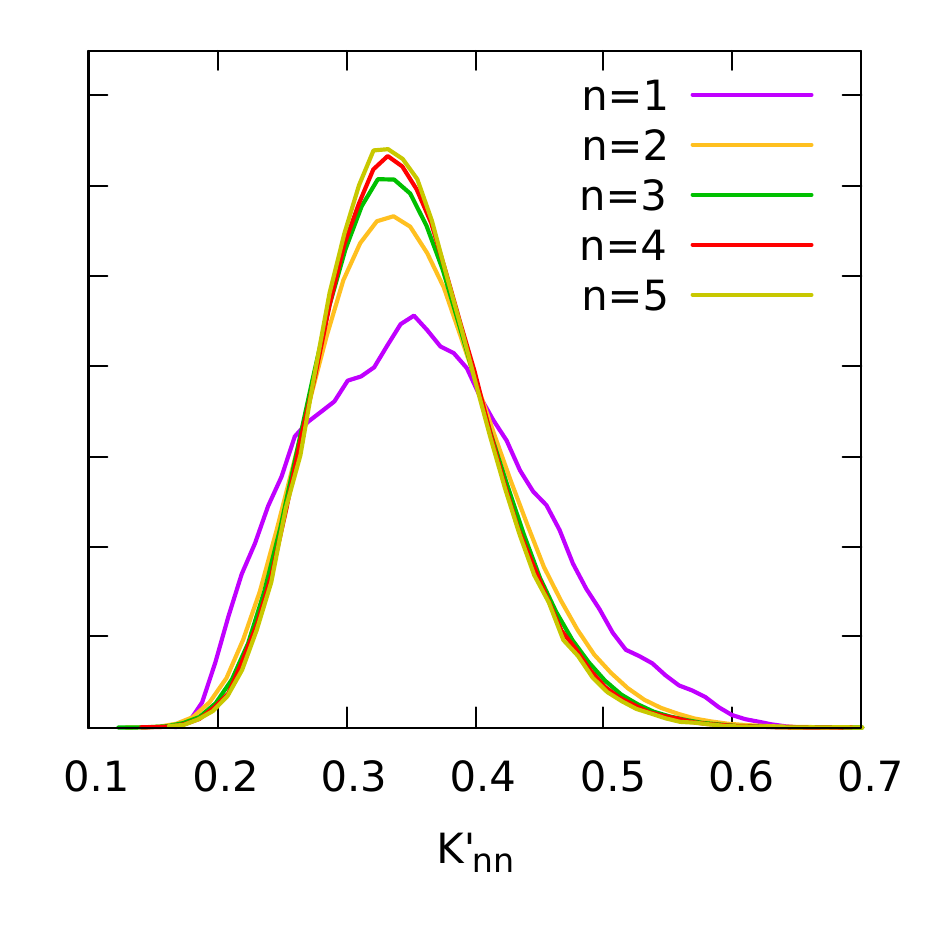}
\end{minipage}%
\begin{minipage}{.155\textwidth}
  \centering
  \includegraphics[scale=0.19]{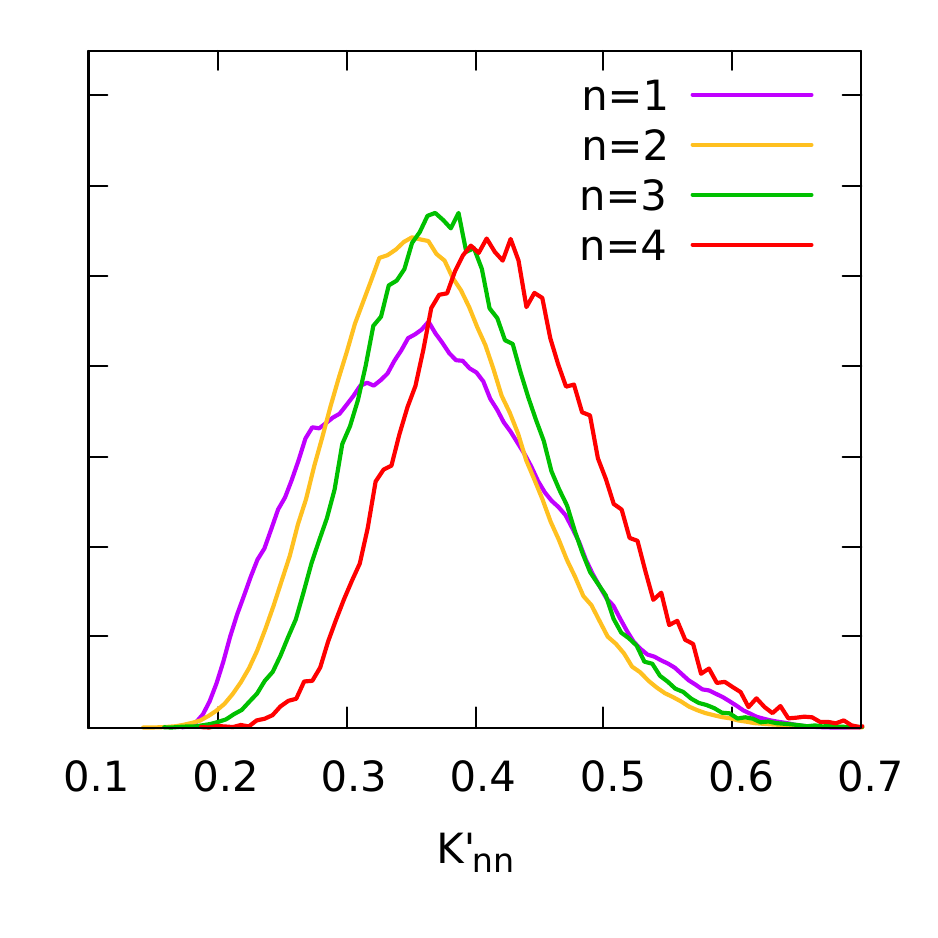}
\end{minipage}
\caption{Distribution of $K'_{\nn}$ for a DIM with $K_{\DI} =  0.60$ (left), 0.609377 (middle), and $0.62$ (right). 
$n$ denotes RG iteration. 
All figures have the same scale.
See the SM \cite{sm} for the optimization details and the number of samples used (for other example models, too).
}
\label{fig:dilute}
\end{figure}

We report in Fig. \ref{fig:dilute} the RG flow of the marginal distribution $Q_{\vec v}(K_{\text{nn}})$. 
The distribution initiating at $K_\DI = K_{\DI,c}$ converges to a fixed distribution, whereas for $K_{\DI}$ less and greater than $K_{\DI,c}$ the distribution approaches the paramagnetic and the ferromagnetic fixed points respectively.   
The marginal distributions $Q_{\vec v} (K_\nnn)$ and $Q_{\vec v}(K_\square)$ show similar behavior \cite{sm}. 
The RG evolution approaches a fixed critical distribution, which has finite width and is non-Gaussian indicating that the 2D DIM remains inhomogeneous at all length scales at criticality.
Thus, the dilute and the pure Ising model in 2D do not share the same fixed-point. 
Indeed, although they have the same critical exponents, according to analytical \cite{Shalaev,andreas,shankar} and numerical \cite{dilute_mc} studies, the singular dependence of the specific heat with respect to temperature is modified by logarithmic factors in the diluted model compared to the pure model \cite{shankar}. 

As detailed in the supplementary material (SM) \cite{sm}, we use 17 coupling functions $U_\beta(\vec K)$ to represent the distribution $P_v(\vec K)$ in the computation of the critical exponents.
With the adopted representation we find a value of $2.018(6)$ for the leading even eigenvalue $\lambda^e$ of the Jacobian matrix, to be compared with $\lambda^e = 2$ of the pure Ising model. 
The error of our estimate was not reduced by adding more $U_\beta$ functions, suggesting that the renormalized Hamiltonian should include more couplings than just nearest neighbor, next nearest neighbor, and square terms for better accuracy. 

\begin{figure}[hth]
\centering
\begin{minipage}{.155\textwidth}
  \centering
  \includegraphics[scale=0.19]{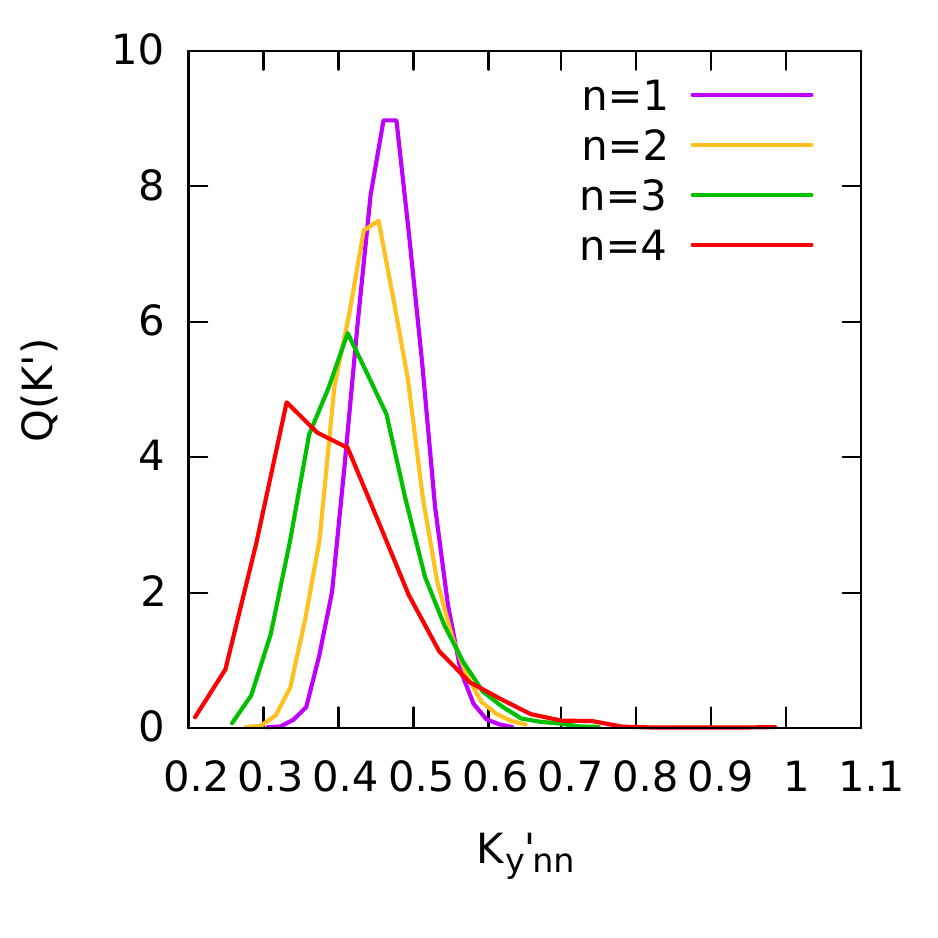}
\end{minipage}%
\begin{minipage}{.155\textwidth}
  \centering
  \includegraphics[scale=0.19]{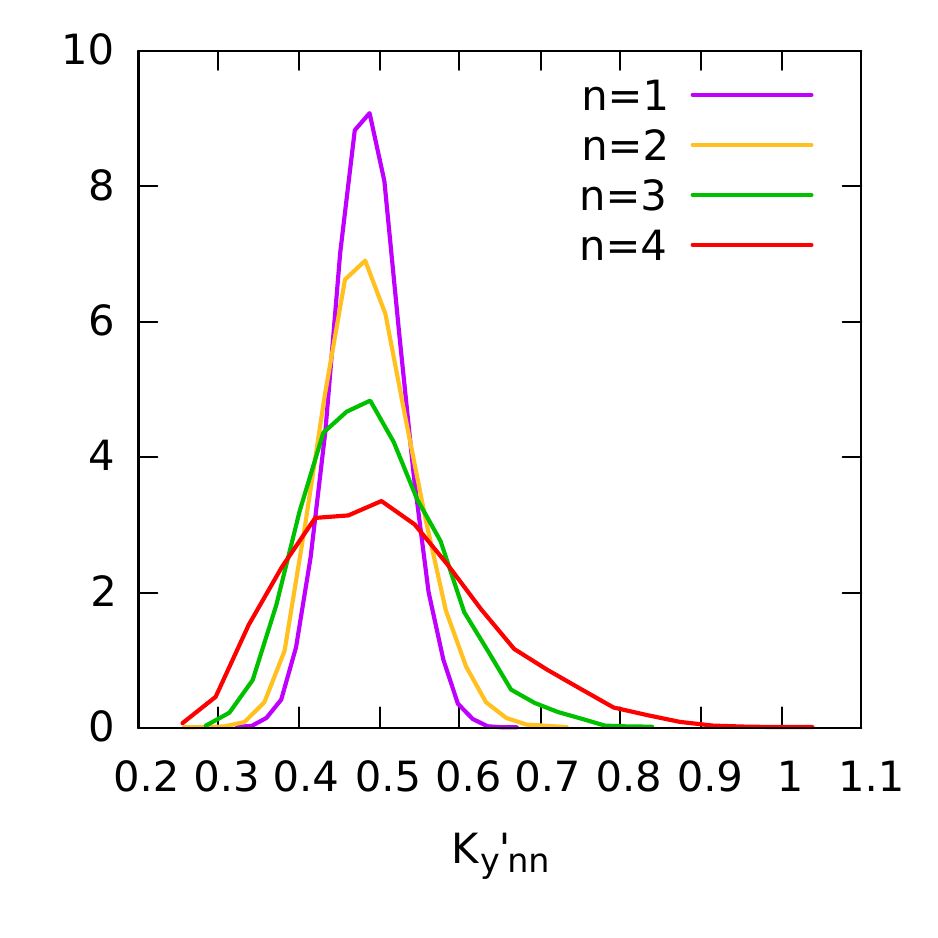}
\end{minipage}%
\begin{minipage}{.155\textwidth}
  \centering
  \includegraphics[scale=0.19]{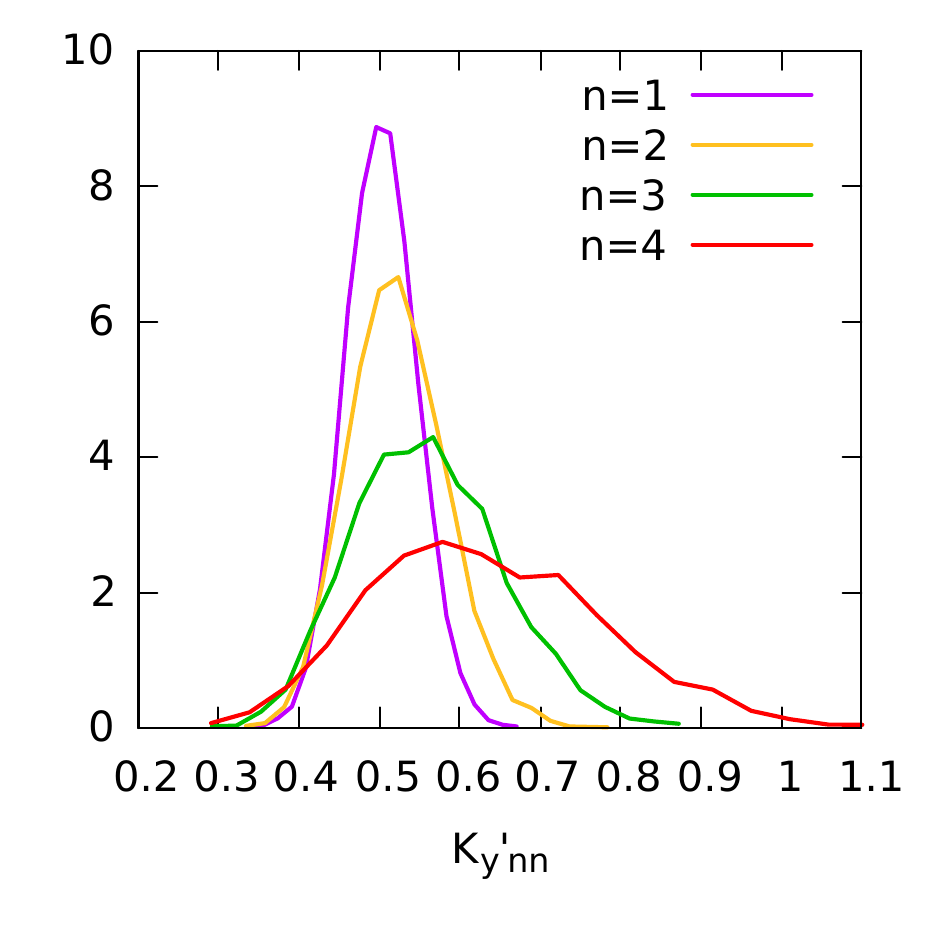}
\end{minipage}
\caption{Distribution of $K_{\nn_y}$ for the Trotter approximation of the TFIM  with $K_{\TFIM} =  0.935$ (left), 1.035 (middle), and $1.135$ (right). 
}
\label{fig:tfim}
\end{figure}

Next we consider the random TFIM on a periodic chain of $L$ spins with Hamiltonian:
\begin{equation}
  \hat{H}_{\TFIM} = - \sum_i k_i \hat{\sigma}^z_i \hat{\sigma}^z_{i+1} - \sum_i h_i \hat{\sigma}^x_i 
  \label{eq:tfim}
\end{equation}
where $\hat{\sigma}^z$ and $\hat{\sigma}^x$ are the Pauli matrices. 
Here $k_i$ and $h_i$ are independently drawn from a Gaussian distribution with standard deviation 0.2, and mean equal to $K_\TFIM$ and $1.0$, respectively. 
By self-duality, the system experiences a ground-state quantum phase transition when $k_i$ and $h_i$ are drawn from the same distribution, i.e. when $K_\TFIM = 1.0$ \cite{fisher_prb}.  
The Trotter-approximation of this model at inverse temperature $\beta$ is an anisotropic nearest-neighbor Ising model on an $L \times \beta m$ periodic 2D lattice with classical Hamiltonian \cite{quantum_mcrg}: 
\begin{equation}
\begin{split}
H_\Trotter &= - \sum_{i=1}^L\sum_{j=1}^{\beta m}  \frac{k_i}{m} \sigma_{i,j} \sigma_{i+1,j} 
\\
&- \sum_{i=1}^L\sum_{j=1}^{\beta m}\frac{1}{2} \ln \left[\coth\left(\frac{h_i}{m}\right)\right] \sigma_{i,j} \sigma_{i,j+1}
\end{split}
\end{equation}
where $\sigma_{i,j}$ is an Ising spin at the $i$th column and $j$th row, and $m$ is the number of Trotter slices.  
As an approximation to $m, \beta \rightarrow \infty$, we use $m = 8, \beta = 16, L = 128$. 
The $2\times 2$ majority-rule block-spin is used despite the anisotropy, as in \cite{quantum_mcrg}.
Four renormalized coupling terms are included in our VMCRG computation: the nearest neighbor coupling in the horizontal ($K_{\nn_x}$) and vertical ($K_{\nn_y}$) directions, the next nearest neighbor $(K_\nnn)$ and smallest plaquette ($K_\square$) couplings. 
In Fig. \ref{fig:tfim}, we report the RG flow of the marginal distribution of $Q(K_{\nn_y})$.  
As in the DIM, both paramagnetic and ferromagnetic fixed-points are discovered. 
At the phase transition, that we found to be at $K_\TFIM = 1.035$ with the adopted Trotter approximation, however, the critical fixed-point is found to have increasing variance, in sharp contrast with the DIM, but consistent with the prediction of SDRG \cite{fisher_prb}.     

We then consider the RFIM in both 2D and 3D:    
\begin{equation}
  H_{\RFIM} = - K_{\RFIM} \sum_{\braket{i, j}} \sigma_i\sigma_j - h_0 \sum_{i}h^i \sigma_i 
\end{equation}
with $K_{\RFIM}$ positive, $\braket{i,j}$ nearest neighbor, and the $h^i$s independent unit Gaussian random variables.  
In both dimensions, we use lattices with linear size $L = 64$ and and adopt the majority rule with $b = 2$ for 3 RG iterations.    
In 2D, we also do a fourth-iteration calculation on a $L=128$ lattice. 

\begin{figure}[hth]
\centering
\begin{minipage}{.25\textwidth}
  \centering
  \includegraphics[scale=0.19]{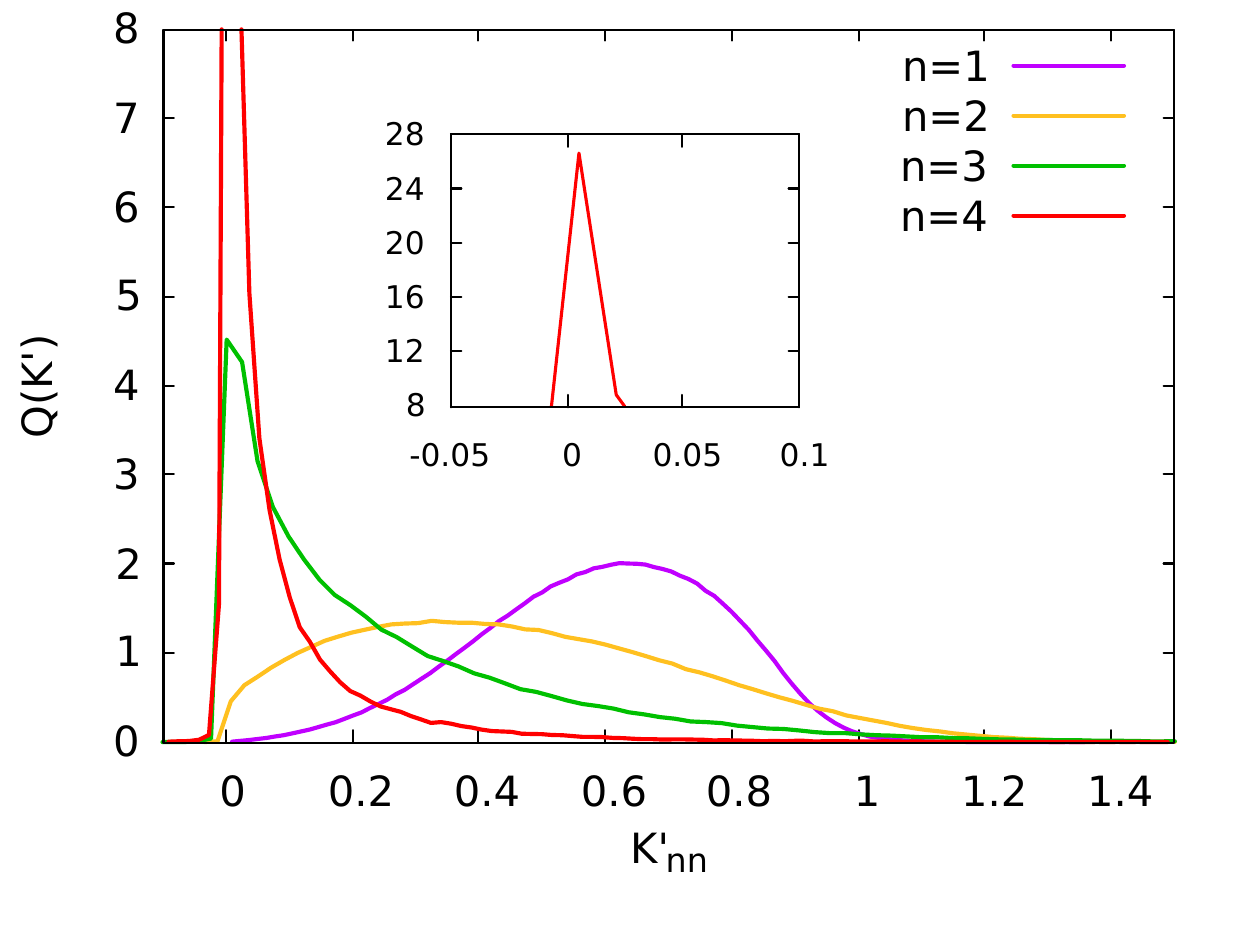}
\end{minipage}%
\begin{minipage}{.25\textwidth}
  \centering
  \includegraphics[scale=0.19]{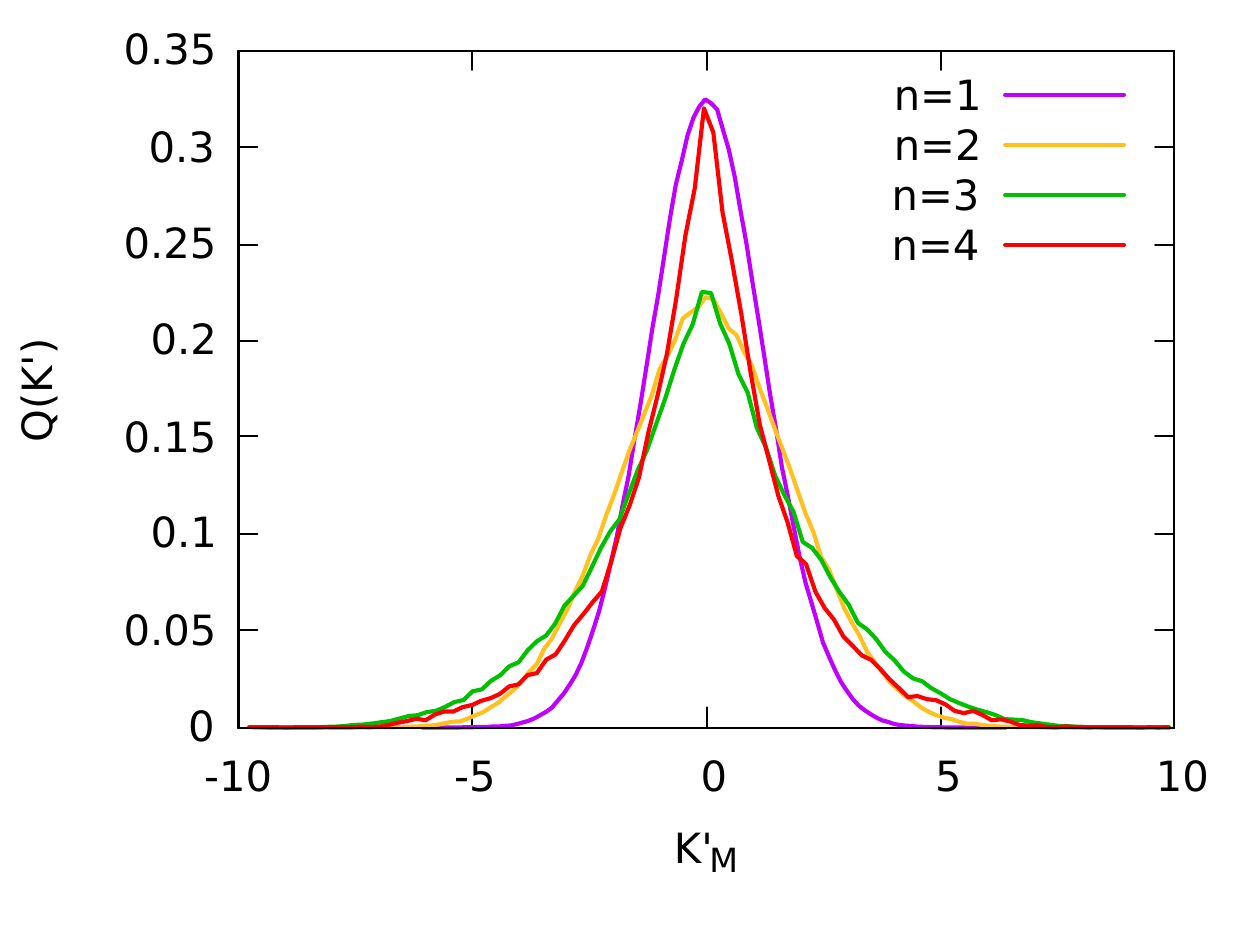}
\end{minipage}
\caption{Distribution of the renormalized nearest neighbor (left) and local magnetization (right) coupling constants for the 2D RFIM with $K_{\RFIM} = 0.8$ and $h_0 = 1$. 
}
\label{fig:sg}
\end{figure}

In the 2D RFIM, we use four couplings, the three even couplings of the DIM and one odd coupling constant $K_{\M}$ describing the strength of the local magnetization $S^{i}_M = \sigma_i$ to account for the random magnetic field. 
As shown in Fig. \ref{fig:sg} (left) when $K_{\RFIM} = 0.8$, a coupling strength well above 0.4407, the critical coupling of the pure Ising model, a random field with strength $h_0 = 1.0$
drives the spin-spin interactions to zero, in accord with the analytical result \cite{aizenman}.  
For the first three iterations, the distribution of $K_\M$ broadens as RG iterates (Fig. \ref{fig:sg}, right), indicating the important role of disorder in suppressing the spin-spin interactions.
When $n = 4$ as the spin-spin interactions have been greatly suppressed, the random fields start to decrease again.  
This must happen, because random fields in an interaction-free spin system renormalize to zero \cite{sm}.   

\begin{figure}[hth]
\centering
\begin{minipage}{.155\textwidth}
  \centering
  \includegraphics[scale=0.19]{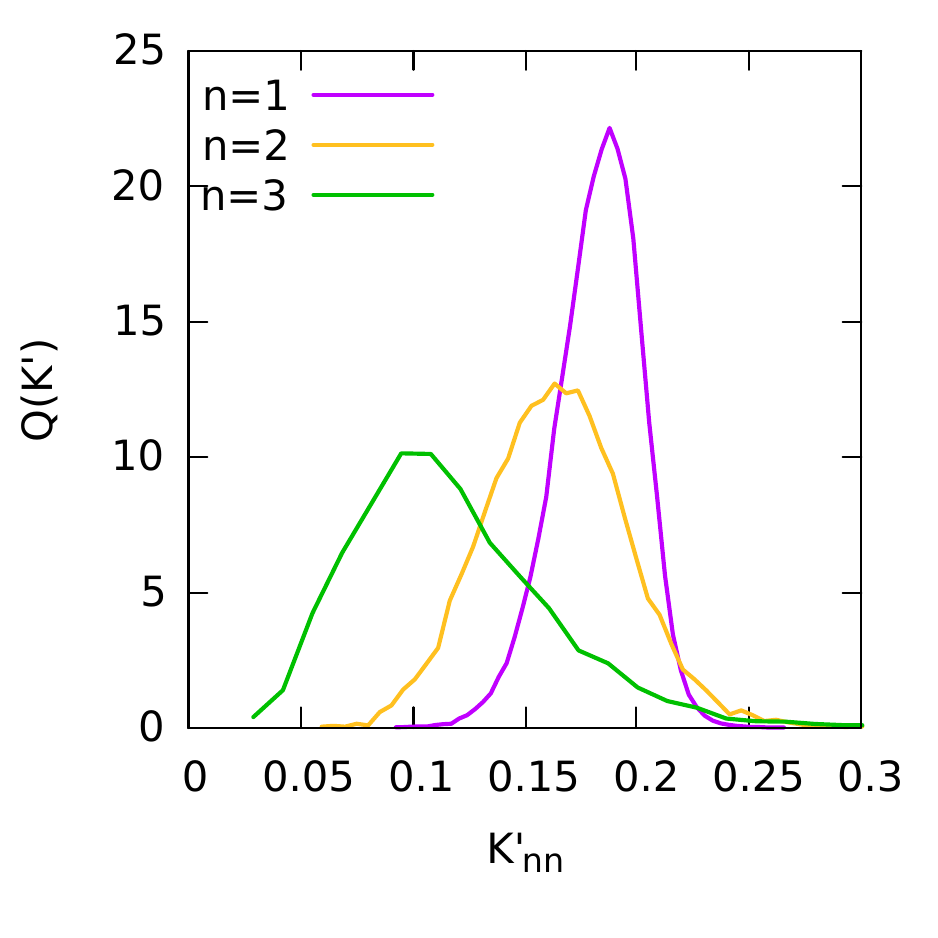}
\end{minipage}%
\begin{minipage}{.155\textwidth}
  \centering
  \includegraphics[scale=0.19]{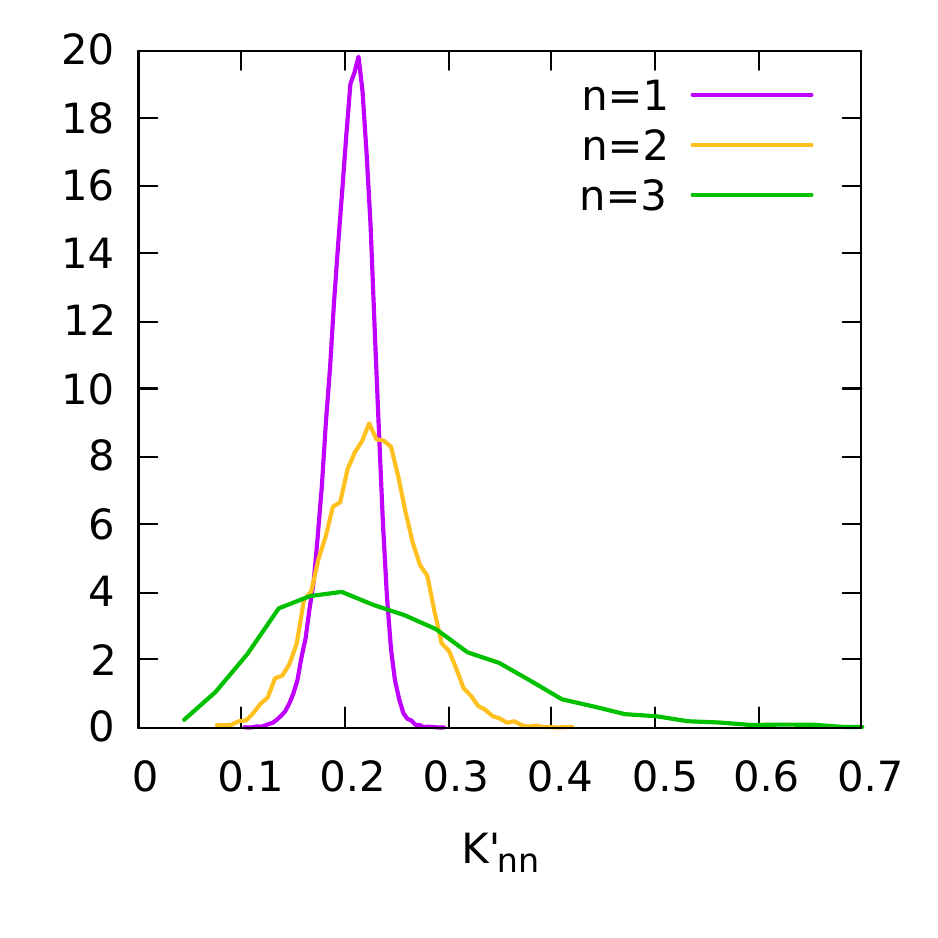}
\end{minipage}%
\begin{minipage}{.155\textwidth}
  \centering
  \includegraphics[scale=0.19]{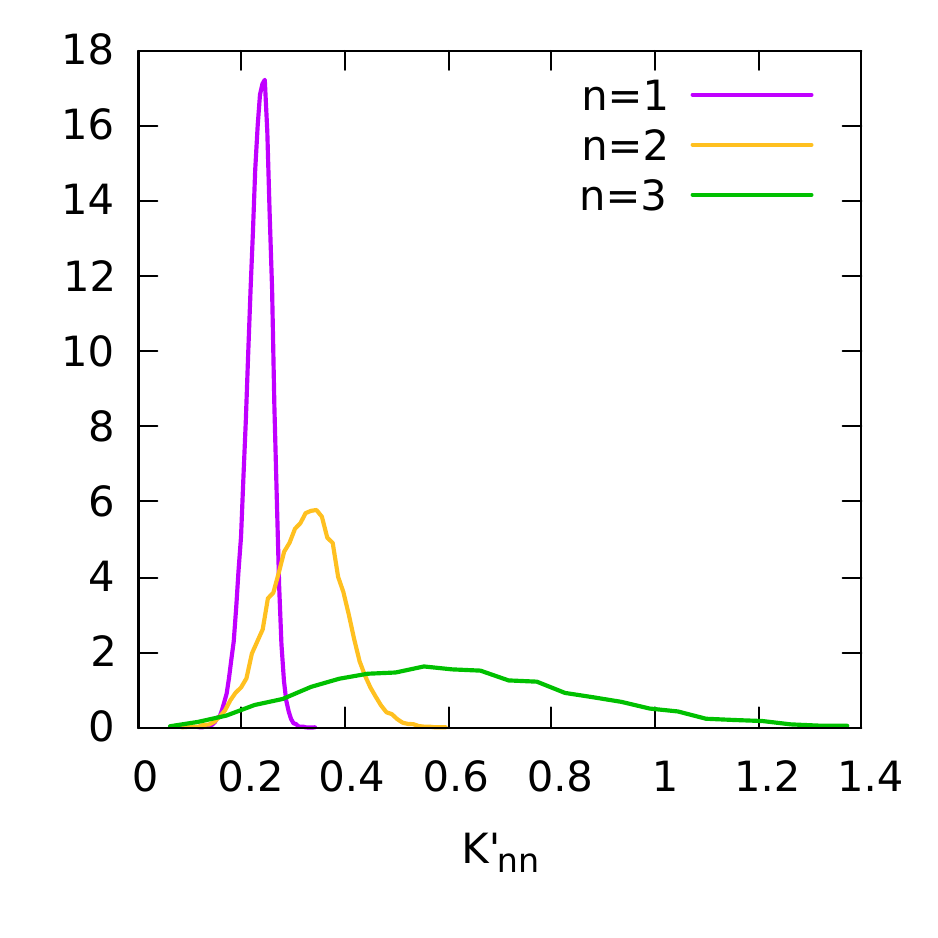}
\end{minipage}
\caption{Distribution of the renormalized nearest neighbor coupling constants for the 3D RFIM with $h_0 = 0.35$, and $K_{\RFIM} = 0.25$ (left), 0.264 (middle), and 0.28 (right). 
}
\label{fig:3D_rfim}
\end{figure}

In the 3D RFIM, we use the six renormalized couplings listed in the SM \cite{sm}, including the nearest neighbor coupling ($K_\nn$) and the local magnetic field ($K_\text{M}$). 
For fixed $h_0 = 0.35$ and varying $K_\RFIM$ the system has been analyzed extensively by finite size scaling for sizes up to $L = 16$, finding that a magnetic transition occurs at $K_\RFIM = 0.2705(3)$ \cite{3D_rfim_MC}.   
We find consistently that the mean value of the Hamiltonian distribution starts drifting toward higher couplings when $K_\RFIM = 0.27$. 
The corresponding variance shows a divergent behavior, suggesting a strong-disorder fixed point. Interestingly, the distribution keeps a fixed non-zero mean with increasing width even below the critical coupling, for $0.26 < K_\RFIM < 0.27$ \cite{sm}. 
Within the number of RG iterations performed, this behavior suggests the presence, in the subcritical region, of magnetic clusters with a disordered distribution of magnetizations. 
The evolution of the renormalized nearest-neighbors coupling is illustrated in Fig. \ref{fig:3D_rfim} for three values of $K_\RFIM$, i.e. $K_\RFIM = 0.25$ (well below the critical coupling), $K_\RFIM = 0.264$ (slightly below the critical coupling), and $K_\RFIM = 0.28$ (well above the critical coupling).      

Finally, we discuss the MC relaxation to equilibrium. 
As shown in the SM \cite{sm}, in the four examples considered the time correlation function of the magnetization in the biased simulation is essentially independent of the lattice size for a given RG iteration when time is measured in MC sweeps, while the relaxation time deteriorates very fast as the lattice sizes get larger in the unbiased ensemble.
We also note that the RG distribution in all the examples above appears to self-average with very little sample-to-sample fluctuation, consistent with the fact that the RG transformation is short-ranged and that the coupling constants are independently distributed before coarse-graining.  

In conclusion, we have described a viable method to computationally realize real-space renormalization for classical statistical systems with quenched disorder. 
Because the biasing mechanism is rather general, the method can be combined with other acceleration schemes, such as cluster algorithms as done for the DIM. 
The method is capable of differentiating systems with a finite-disorder fixed point, like the DIM, and systems with a strong-disorder fixed point, like the disordered Ising chain in transverse field and the 3D RFIM. 
In the former case the method allows us to compute the critical coupling and the critical exponents. 
In the latter case, the method can give new insights in the behavior of the disordered distribution close to criticality. 

\begin{acknowledgments}
All the codes used in this project were written in C\texttt{++}, and will be available upon request. We acknowledge support from DOE Award DE-SC0017865.  
\end{acknowledgments}

\bibliographystyle{apsrev}
\bibliography{abc}
\end{document}